\documentclass[a4paper,12pt]{article}

\newcommand{\hepth}[1]
      {{\tt hep-th/#1}}

\begin{document}

\begin{titlepage}

\begin{flushright}
UUITP-09/00\\
hep-th/0011054
\end{flushright}

\vspace{1cm}

\begin{center}
{\huge\bf  Wound String Scattering \\[5mm]
           in NCOS Theory}
\end{center}
\vspace{5mm}

\begin{center}

{\large 
Fredric Kristiansson and 
Peter Rajan} \\

\vspace{5mm}

Institutionen f\"or Teoretisk Fysik, Box 803, SE-751 08
Uppsala, Sweden

\vspace{5mm}

{\tt
fredric.kristiansson@teorfys.uu.se, \\
peter.rajan@teorfys.uu.se
}

\end{center}

\vspace{5mm}

\begin{center}
{\large \bf Abstract}
\end{center}
\noindent

We calculate the amplitude for a non-excited closed string
with nonzero winding number to scatter from a D-string with a near
critical $E$ field. We go to the NCOS limit and observe that we get
the same result if we adopt another approach put forward by
Gomis and Ooguri.

\vfill
\begin{flushleft}
November 2000
\end{flushleft}
\end{titlepage}
\newpage

\section{Introduction}

\bigskip

There has recently been a lot of interest in the non-commutative open
string (NCOS) theory living on a D-string \cite{sst,ncos,om}. The NCOS
limit is reached by tuning the electric field along the D-string to
its critical value which causes an infinite rescaling of the string
tension and coupling. It is now understood that the NCOS limit is
meaningful even in the absence of a D-brane, as demonstrated in the
papers \cite{dgk, gooo}. This more general theory is called wound
string theory \cite{dgk}, or non-relativistic closed string theory
\cite{gooo}. A number of calculations have been performed in the NCOS
theory including various scattering amplitudes \cite{ncos,km,hk} and
supergravity duals \cite{harm,sah}. As far as closed strings are
concerned, NCOS theory has the usual rules for computing scattering
amplitudes.

One can approach NCOS theory from a different direction using the set
of rules proposed by \cite{gooo}.  These rules
are obtained by taking a low energy limit already in the action,
something which appears to precisely reproduce NCOS theory if a
D-brane is present. In this sense the NCOS limit
can be thought of as a non-relativistic limit.

It is certainly of value to confirm that these two different
approaches give equivalent results. In this note we calculate the
amplitude for a non-excited closed string to scatter from a
D-string. In section 2 we use standard techniques for reaching the
amplitude, taking the NCOS limit only afterwards. The strings have
non-zero winding numbers and we will see that there are restrictions
on the winding for the kinematic invariants to remain finite in the limit.
The corresponding calculation with the new set of rules is shown in
section 3, which, as expected, gives the same scattering amplitude.
Finally, in section 4 we present some conclusions.

\section{NCOS scattering}

\bigskip

We calculate the amplitude for a non-excited NS-NS
closed string with nonzero winding number to scatter from a D-string with a
near critical $E$ field. Similar calculations for unwound strings
are performed in \cite{gm,hashk} (without $E$-field) and \cite{gkp} 
(with a near critical $E$-field). Taking the
D-string along the $x_1$ direction, we split the metric according to
\begin{equation}
g_{ab}=\eta_{ab}, \qquad g_{ij}=h\delta_{ij}, \qquad a,b=0,1, \qquad
i,j=2,\ldots,9, 
\end{equation}
and define
the `image method'
matrix ${D^{\mu}}_{\nu}$ as \cite{gkp},
\begin{equation}
{D^{a}}_{b}=\frac{1}{1-E^{2}}\left( 
\begin{array}{cc}
1+E^{2} & 2E \\ 
2E & 1+E^{2}
\end{array}
\right) ,\qquad {D^{i}}_{j}=-\delta _{ij}.  \label{D}
\end{equation}
Let $p_{\mu}=(k_0,k_1+wR/\alpha',k)$ and
$\tilde{p}_{\mu}=(k_0,k_1-wR/\alpha',k)$ denote the momenta
of the left- and right moving modes respectively, where $w$ 
is the winding number around the
compact $x_1$ direction. For simplicity we choose $k_1=n/R=0$.
We look at states polarized
transversely to the D-string, i.e.~each polarization tensor
$\epsilon_{\mu\nu}$ obeys $\epsilon_{ab}=0$. Additionally we require
that $\epsilon^{T}=\epsilon$ and $\mbox{Tr}\epsilon=0$.
We may now write the scattering amplitude as \cite{gm,gkp}
\begin{eqnarray} 
& & A \sim \frac1{\lambda} \sqrt{1-E^2}\int d^2z_1 d^2z_2 
\epsilon_{1\mu\nu}
\epsilon_{2\sigma\kappa} \nonumber \\ 
\label{Eq:amp1}
& & \qquad \qquad \qquad \times \ \left\langle 
V_{-1}^{\mu}(p_{(1)},z_1)V_{-1}^{\nu}(D\cdot\tilde{p}_{(1)},\bar{z}_1) 
V_0^{\sigma}(p_{(2)},z_2)V_{0}^{\kappa}(D\cdot\tilde{p}_{(2)},\bar{z}_2)
\right\rangle,
\end{eqnarray} 
where the vertex operators are given by
\begin{eqnarray}
V_{-1}^{\mu}(p,z) & = & e^{-\phi(z)} \psi^{\mu}(z) e^{ip\cdot X(z)}, 
\nonumber \\
V_0^{\mu}(p,z) & = & (\partial X^{\mu}(z)+ip\cdot \psi(z)\psi^{\mu}(z))
e^{ip\cdot X(z)}.
\end{eqnarray} 
In (\ref{Eq:amp1}) we used the fact that for transverse polarizations
we get $\epsilon\cdot D=D\cdot\epsilon=-\epsilon$. Keeping track of the
different modes it is seen that the results can be inferred from
\cite{gm,gkp} by simply replacing $D\cdot p$ with
$D\cdot\tilde{p}$. Momentum conservation in the directions
parallel to the D-string requires that
\begin{equation}\label{eq:mc} 
p_{(1)} + D\cdot\tilde{p}_{(1)} + p_{(2)} + D\cdot\tilde{p}_{(2)} = 0,
\end{equation}
which implies
\begin{equation}\label{eq:q}
\begin{array}{c}
p_{(1)}\cdot p_{(2)} = \tilde{p}_{(1)}\cdot \tilde{p}_{(2)} = 
2t/\alpha', \qquad 
p_{(1)}\cdot D\cdot \tilde{p}_{(2)} = \tilde{p}_{(1)}\cdot D \cdot p_{(2)} 
 \\[3mm]
\mbox{and} \qquad p_{(1)}\cdot  D
\cdot\tilde{p}_{(1)}=p_{(2)}\cdot  D \cdot\tilde{p}_{(2)}=2s/\alpha',
\end{array}
\end{equation}
where we have defined the kinetic invariants $s$ and $t$. 
Multiplying (\ref{eq:mc}) by $\epsilon_i$ we see that $p_{(i)}\cdot
\epsilon_j=\tilde{p}_{(i)} \cdot \epsilon_j$. Using these relations, all
but one of the kinematic factors vanish or cancel out, and what remains is
identical to the corresponding result in \cite{gm} except for a change
$p_{(2)}\cdot D \cdot p_{(2)} \mbox{Tr}(\epsilon_1\cdot
\epsilon_2)\rightarrow
p_{(2)}\cdot D \cdot \tilde{p}_{(2)} \mbox{Tr}(\epsilon_1\cdot \epsilon_2)$.
Below we show a few steps in the calculation. Eliminating terms which
eventually add up to zero, we can write
\begin{eqnarray}
A & \sim & \frac1{\lambda} \sqrt{1-E^2}\int d^2z_1 \, d^2z_2 
 C(z_1,\bar{z}_1,z_2,\bar{z}_2) 
\nonumber \\
& & \qquad \qquad \times \ \left\langle 
:e^{ip_{(1)}\cdot X(z_1)}::e^{iD \cdot \tilde{p}_{(1)}\cdot X(\bar{z}_1)}:
:e^{ip_{(2)}\cdot X(z_2)}::e^{iD \cdot \tilde{p}_{(2)}\cdot X(\bar{z}_2)}:
\right\rangle,
\end{eqnarray}
where $C(z_1,\bar{z}_1,z_2,\bar{z}_2)$ contains the contractions 
not involving the $X^\mu(z)$'s.  Fixing
the $SL(2,R)$ invariance by putting $z_1=i$ and $z_2=iy$ we get the
Jacobian $d^2z_1d^2z_2 \rightarrow 4(1-y^2)dy$ and, for the
contractions
\begin{equation}
C(i,-i,iy,-iy) = \frac{1}{(1-y^2)^2}
\mbox{Tr}(\epsilon_1\cdot\epsilon_2) \ s
\end{equation}
and
\begin{eqnarray}
& &
\left\langle 
:e^{ip_{(1)}\cdot X(i)}::e^{iD \cdot \tilde{p}_{(1)}\cdot X(-i)}:
:e^{ip_{(2)}\cdot X(iy)}::e^{iD \cdot \tilde{p}_{(2)}\cdot X(-iy)}:
\right\rangle \ = \nonumber \\
& & \qquad \qquad \qquad \qquad \qquad \qquad \qquad 
= \  (-1)^s 2^{s}(2y)^{s}(1-y)^{2t}(1+y)^{-2s-2t}.
\end{eqnarray}
A change of variables to $y=(1-\sqrt{x})/(1+\sqrt{x})$ gives a
standard Euler beta integral, and we get 
\begin{equation}
 A =  \frac1{\lambda}\sqrt{1-E^2}\frac{\Gamma(t)\Gamma(s+1)}
{\Gamma(1+s+t)}\mbox{Tr}(\epsilon_1\cdot\epsilon_2)\ s.
\end{equation}
Writing out $s$ explicitly we have
\begin{equation}
s = -\frac{1+E^2}{1-E^2} \frac{\alpha'}{2} (k_0)^2 
- \frac{1+E^2}{1-E^2}\frac{\alpha'}{2} \left( \frac{wR}{\alpha'} \right)^2 
- \frac{\alpha'}{2h} k^2 - \frac{2EwR}{1-E^2}k_0.  \label{s}
\end{equation}
Notice
that the mass-shell condition (for arbitrary closed string states) implies
that 
\begin{equation} \label{k0}
k_{0}=\sqrt{\left(\frac{wR}{\alpha'}\right)^{2}
+\left(\frac{n}{R}\right)^{2}+\frac{k^{2}}{h}
+\frac{2}{\alpha'}(N+\tilde{N})}~.
\end{equation}
We will be looking at the special case were $N=\tilde{N}=n=0$. 
Using (\ref{k0}) to eliminate $k_0$ in (\ref{s}), we conclude that 
\begin{equation}
s=-\frac{\alpha'}{1-E^2} \left(\frac{EwR}{\alpha'} + 
   \sqrt{\left(\frac{wR}{\alpha'}\right)^2+\frac{k^2}{h}}\right)^2.
\end{equation}
We now take the NCOS limit $\varepsilon \rightarrow 0$ with
$E=1-\varepsilon/2$, $\alpha'=\varepsilon \alpha_e'$ and
$h=\varepsilon$. For $w>0$, which is needed for $s$ to remain finite in
the limit, we get as $\varepsilon \rightarrow 0$
\begin{equation}
s=-\alpha'_e\left(\frac{wR}{2\alpha'_e} + 
\frac{\alpha'_e k^2}{2wR}\right)^2=-\alpha'_e(k_0^{\mathrm{NCOS}})^2,
\end{equation}
where we have introduced $k_0^{\mathrm{NCOS}}$ as an abbreviation.
Note that in the absence of an electric field we can write the ``image
method'' matrix as $D=V-N$ where $V$ and $N$ projects out parallel and
perpendicular (to the brane) parts respectively so that
\begin{equation}
s=\frac{\alpha'}{2}(pD\tilde{p}) = \frac{\alpha'}{4}(p+D\tilde{p})^2 =
\frac{\alpha'}{4}(V(p+\tilde{p}))^2=\alpha' k_0^2 g^{00}.
\end{equation}
The net effect of the (near critical) electric field is in the
present case $\alpha'g^{\mu\nu} \rightarrow \alpha'_e
\eta^{\mu\nu}$.  Taking the limit amounts to replace $k_0$ with
$k_0^{\mathrm{NCOS}}$. Analogously, for $t$ we have
\begin{equation}
t=\frac{\alpha'_e}{4}\left(k_{(1)}²+k_{(2)}²+2k_{(1)}k_{(2)}\right),
\end{equation}
which is just $\alpha'g^{\mu\nu} \rightarrow \alpha'_e \eta^{\mu\nu}$
applied to
\begin{eqnarray}
t & = & \frac{\alpha'}{4}\left(p_{(1)}+p_{(2)}\right)^2 
  \ = \ \frac{\alpha'}{4}\left(V(p_{(1)}+p_{(2)})
               +N(p_{(1)}+p_{(2)})\right)^2
\nonumber
\\
  & = & \frac{\alpha'}{4}\left(k_{(1)} + 
            k_{(2)}\right)_i g^{ij}\left(k_{(1)}+k_{(2)}\right)_j.
\end{eqnarray}

\section{Wound string theory calculation}

\bigskip

It is interesting to check whether we arrive at the same result in
wound string theory using the set of rules proposed by \cite{gooo}.
These rules are obtained by taking the critical limit already in the action.
In this limit, the bosonic part of the world sheet action is \cite{gooo}
\begin{equation}
S=\int \frac{d^2z}{2\pi}\left(\beta\bar{\partial}\gamma
+\bar{\beta}\partial\bar{\gamma}+\frac{1}{4\alpha'_e}\partial\gamma\bar{\partial}\bar{\gamma}+\frac{1}{\alpha'_e}\partial
X^i \bar{\partial}X_i \right),
\end{equation}
with $\gamma=X^0+X^1,\;\bar{\gamma}=-X^0+X^1$ and the $\beta$'s
as Lagrange multipliers.  The zero mode parts of the momenta conjugate
to $\gamma$ and $\bar{\gamma}$ are
\begin{eqnarray}
\frac{1}{2}(k_0+k_1)=i\beta_0+\frac{wR}{4\alpha'_e} & \quad
\mbox{and} \quad  & 
\frac{1}{2}(k_0-k_1)=i\bar{\beta}_0+\frac{wR}{4\alpha'_e}
\end{eqnarray}
and the Virasoro constraint is
\begin{eqnarray}
i\beta_0=\frac{N}{wR}+\frac{\alpha'_e k^2}{4wR}, & \qquad & 
i\bar{\beta}_0=\frac{\tilde{N}}{wR}+\frac{\alpha'_e k^2}{4wR}.
\end{eqnarray}
Thus the energy in wound string theory, $k_0^{\mathrm{W}}$, is given by
\begin{equation}
k_0^{\mathrm{W}}=v+\bar{v}+\frac{wR}{2 \alpha'_e } = 
\frac{wR}{2 \alpha'_e}+\frac{\alpha'_e k²}{2wR}+\frac{N+\tilde{N}}{2wR},
\end{equation}
where $v$ and $\bar{v}$ are the eigenvalues of zero mode
$i\bar{\beta}$ and $i\beta$ respectively. Note that 
$k_0^{\mathrm{W}}=k_0^{\mathrm{NCOS}}$.
The momentum $k_1$ conjugate to the compact direction $x_1$ is given by
\begin{equation}
k_1=\bar{v}-v.
\end{equation}
For $N+\tilde{N}=0$ and $k_1=0$ we get
\begin{equation}
v=\bar{v}=\frac{\alpha'_e k²}{4wR}.
\end{equation}
We would now like to check that the scattering amplitude calculated in
this framework agrees with the NCOS amplitude in the previous section.
To do this, we make use of the propagators below, where $\mu,\nu\ge
2$,
\begin{eqnarray}
\big\langle X^\mu(z_i)X^\nu (z_j)\big\rangle = 
-\frac{\alpha'_e}{2}\eta^{\mu\nu}\log(z_i-z_j),
& & 
\big\langle X^\mu (z_i)\tilde{X}^\nu (\bar{z}_j)\big\rangle =
\frac{\alpha'_e}{2}\eta^{\mu\nu}\log(z_i-\bar{z}_j),
\nonumber 
\\
\big\langle \bar{\gamma}(\bar{z}_i)\ {\textstyle
\int}^{\bar{z}_j}\bar{\beta} 
\big\rangle
=\log(\bar{z}_i-\bar{z}_j),
& & 
\big\langle \gamma(z_i)\ {\textstyle \int}^{z_j}\beta \big\rangle 
=\log(z_i-z_j),
\nonumber 
\\
\big\langle \gamma(z_i)\bar{\gamma}(\bar{z}_j) \big\rangle 
=4\alpha'_e\log(z_i-\bar{z}_j),
& &
\big\langle {\textstyle \int}^{z_i}\beta\ 
{\textstyle \int}^{\bar{z}_j}\bar{\beta}\big\rangle
=\frac{1}{4\alpha'_e}\log(z_i-\bar{z}_j).
\end{eqnarray}
The interesting part of the amplitude, $\big\langle 
\prod :e^{ip_{(i)}X}:\big\rangle$, is now given by
\begin{eqnarray}
&&
\big\langle :e^{iv_{(1)}\gamma(z_1)+
   iw_{(1)}R\int^{z_1}\beta+ik_{(1)}X(z_1)}:
:e^{iv_{(1)}\bar{\gamma}(\bar{z}_1)+
   iw_{(1)}R\int^{\bar{z}_1}\bar{\beta}+ik_{(1)} \tilde{X}(\bar{z}_1)}:
\nonumber
\\ 
&&
\qquad \qquad :e^{iv_{(2)}\gamma(z_2)+
   iw_{(2)}R\int^{z_2}\beta+ik_{(2)}X(z_2)}:
:e^{iv_{(2)}\bar{\gamma}(\bar{z}_2)+
   iw_{(2)}R\int^{\bar{z}_2}\bar{\beta}+ik_{(2)}\tilde{X}(\bar{z}_2)}:
\big\rangle \ = \nonumber 
\\ 
&&
\qquad = \  
(z_1 - \bar{z}_1)^{-\frac{\alpha'_e}{4}(\frac{wR}{\alpha'_e}+4v)^2}
(z_2 - \bar{z}_2)^{-\frac{\alpha'_e}{4}(\frac{wR}{\alpha'_e}+4v)^2}
\nonumber
\\ 
&& 
\qquad \qquad \qquad 
\times \ |z_1-z_2|^{4vwR + \alpha'_e k_{(1)} k_{(2)}}
|z_1-\bar{z}_2|^{\frac{\alpha'_e}{2}(\frac{wR}{\alpha'_e}+4v)^2 
       -4vwR - \alpha'_e k_{(1)} k_{(2)} }
\nonumber 
\\
&&
\qquad = \ 
(z_1-\bar{z}_1)^s (z_2-\bar{z}_2)^s
|z_1-z_2|^{2t}|z_1-\bar{z}_2|^{-2s-2t}
\nonumber 
\\
&&
\qquad = \ 
(-1)^s 2^s (2y)^s (1-y)^{2t}(1+y)^{-2t-2s}.
\end{eqnarray} 
In the second equality we have used that $s$ and $t$ can be written as
\begin{eqnarray}
s&&=-\frac{\alpha'_e}{4}\left(\frac{wR}{\alpha'_e}+4v\right)^2 \\
t&&=-v_{(1)} w_{(2)} R - v_{(2)} w_{(1)} R + \frac{\alpha'_e}{2}k_{(1)} k_{(2)}=2vwR+\frac{\alpha'_e}{2}k_{(1)} k_{(2)}\nonumber,
\end{eqnarray}
where $ w=w_{(1)} =-w_{(2)} $ and $ v = v_{(1)}= -v_{(2)} $.  As
expected, we recover the NCOS amplitude as far as the bosonic fields
are concerned. The fermionic fields can be treated analogously, adding
up to an amplitude which is identical to that in the previous section.

\section{Conclusions}

In this paper we have calculated the amplitude for a non-excited closed
string to scatter from a near critical D-string. We allow for nonzero
winding numbers so that the kinetic invariants can remain finite in the
limit. Two different approaches to the calculation were presented. We
started off by using standard rules, waiting until the end before
taking the NCOS limit.  Then we adopted the set of rules put
forward in \cite{gooo}, which amounts to taking a proper limit already
in the action. We have thus been able to confirm that the two approaches
give rise to equivalent amplitudes.

\section*{Acknowledgements}

We would like to thank Ulf Danielsson and Martin Kruczenski for their
invaluable support. We are also grateful to Alberto G\"uijosa.

\end{document}